\documentclass[12pt]{article}
\usepackage{amssymb}
\usepackage{cite}

\def\courbure{{\cal R}}

\newcommand{\nc}{\newcommand}
\nc{\beq}{\begin{equation}}
\nc{\eeq}{\end{equation}}
\nc{\beqa}{\begin{eqnarray}}
\nc{\eeqa}{\end{eqnarray}}
\nc{\lra}{\leftrightarrow}

\nc{\sss}{\scriptscriptstyle}
{\nc{\lsim}{\mbox{\raisebox{-.6ex}{~$\stackrel{<}{\sim}$~}}}
{\nc{\gsim}{\mbox{\raisebox{-.6ex}{~$\stackrel{>}{\sim}$~}}}

\def\yT{{y_{\sss T}}}
\def\yTp{{y_{\sss T'}}}
\def\yP{{y_{\sss P}}}
\def\kT{{k_{\sss T}}}
\def\kTp{{k_{\sss T'}}}
\def\kP{{k_{\sss P}}}
\def\sT{{\sigma_{\sss T}}}
\def\sTp{{\sigma_{\sss T'}}}
\def\sP{{\sigma_{\sss P}}}
\def\LT{{\Lambda_{\sss T}}}
\def\LTp{{\Lambda_{\sss T'}}}
\def\LP{{\Lambda_{\sss P}}}
\def\VT{{V_{\sss T}}}
\def\VTp{{V_{\sss T'}}}
\def\VP{{V_{\sss P}}}

\begin{document}

\begin{titlepage}
\begin{flushright}
McGill 99-30\\
LBNL 44307\\
UCB-PTH-99/45\\
Saclay t99/108 \\
hep-ph/9909496 \\
\end{flushright}

\vskip.5cm
\begin{center}
{\huge{\bf Inflating Intersecting Branes\\
 and Remarks\\ \vskip0.4cm
 on the Hierarchy Problem}}
\end{center}
\vskip1.5cm

\centerline{ J. Cline$^{a,*}$, C. Grojean$^{b,c,\dagger}$ 
and G. Servant$^{a,d,\star}$ }
\vskip 15pt
\centerline{$^{a}$ Physics Department, McGill University,
Montr\'eal, Qu\'ebec, Canada H3A 2T8}
\vskip 3pt
\centerline{$^{b}$ Department of Physics, 
University of California, Berkeley, CA 94720}
\vskip 3pt
\centerline{$^{c}$ Theoretical Physics Group,
 Lawrence Berkeley National Laboratory, Berkeley, CA 94720}
\vskip 3pt
\centerline{$^{d}$ CEA-SACLAY, Service de Physique Th\'eorique, 
F-91191 Gif-sur-Yvette, France}
\vglue .5truecm

\begin{abstract}
\vskip 3pt

We generalize solutions of Einstein's equations for intersecting branes in
higher dimensional spacetimes to the nonstatic case, modeling an expanding
universe.  The relation between the Hubble rate, the brane tensions, and
the bulk cosmological constant is similar to the case of a single 3-brane
in a 5-dimensional spacetime.  However, because the bulk inflates as well
as the branes, this class of solutions suffers from Newton's constant
tending toward zero on the TeV brane, where the Randall-Sundrum mechanism
should solve the weak scale hierarchy problem.  The strength of gravity
remains constant on the Planck brane, however.

\end{abstract}

\vfill
\leftline{* jcline@physics.mcgill.ca} 
\leftline{$\dagger$ CMGrojean@lbl.gov}
\leftline{$\star$ servant@spht.saclay.cea.fr}

\end{titlepage}

Although the possibility of extra spatial dimensions is an old idea, it
has received more attention lately because of a new twist: perhaps our
universe looks 4-dimensional not because of the smallness of the extra
dimensions, but because we are trapped on a 3D slice (a 3-brane)
\cite{Trapping,LargeDim,RS}.  This simple variation has created a wealth of
potential new physics signals and hints of solutions to long-standing
puzzles.

At first sight, the brane universe scenario poses a new problem: if one
embeds a 3-brane with tension (energy density) $\sigma$ in an empty
$(4+1)$-D spacetime, the space in the 3-brane inflates with a Hubble
constant given by \cite{LOW,BDL,Kaloper,Nihei}
\beq
\label{wrongrate}
	H^2 = \left({\kappa^2\sigma\over 6}\right)^2
\eeq
Here $\kappa^2$ is the analog of $8\pi G_N = M_P^{-2}$ (where $M_P$ is the
Planck mass) in $(4+1)$-D gravity.  The linear dependence $H\sim\sigma$
is contrary to the usual Friedmann equation which gives
$H\sim\sqrt{\sigma}$.  If one tries to model normal cosmology on such a
brane universe by replacing the constant tension $\sigma$ with an energy
density $\rho$ which decreases with the expansion, then $H$ varies with
time and is given by $1/4t$ instead of $1/2t$ in a radiation dominated
era.  This corresponds to a scale factor growing like $t^{1/4}$ rather
than the usual $t^{1/2}$.  It is likely that such a radical change to the
expansion rate can be ruled out using primordial big bang nucleosynthesis
\cite{BDL}. 

However, one can to a good approximation recover the usual rate of
expansion by keeping a nonzero value for the constant part of the brane
tension, and canceling its inflationary effect by adding a negative
cosmological constant $\Lambda_b$ in the bulk (the full 5 spatial
dimensions) \cite{CGKT,CGS}.  Letting $\rho$ denote the
time-varying part of the energy density on the brane,
eq.\ (\ref{wrongrate}) is modified to read
\beq
\label{rightrate}
    H^2 = {\kappa^4(\sigma+\rho)^2\over 36} + {\kappa^2\Lambda_b\over 6}.
\eeq
By tuning the value of $\Lambda_b$ to 
\beq
\label{RSrel}
	\Lambda_b = -\kappa^2\sigma^2/6,
\eeq
the quadratic term in $\Lambda$ gets canceled, so that the universe is
static when $\rho=0$, as expected.  Further tuning $\sigma$ to the value
\beq
\label{Lrel}
	\sigma = {6\over\kappa^4 M_P^2},
\eeq
one finds that the leading correction to $H^2$ for $\rho\ll \sigma$
agrees exactly with the usual Friedmann equation.  Only for
$\rho\gsim\sigma$ does the unusual $H\sim\rho$ behavior start to
reappear.

In addition to solving the problem of cosmological expansion, the
relation (\ref{RSrel}) has another possible benefit: it might afford a
solution to the hierarchy problem, {\it i.e.}, the question of why the
weak scale $M_W$ is so much smaller than $M_P$
\cite{RS,Gogberashvili,RL}\footnote{The question of the gauge coupling
unification has also been addressed in \cite{RSunification}}.
Randall and Sundrum
noticed that the solution for the metric on the 4D space is
exponentially suppressed away from the 3-brane, in the direction $y$ of
the 5th dimension:
\beqa
	ds^2 &=& a^2(y) \left( - d\tau^2 + \sum_{i=1}^3 dx_i^2\right) 
	+ b^2 dy^2;\nonumber\\
	a(y) &=& \exp(- \kappa^2 b \Lambda|y|/6)
\eeqa

At a given distance $\yT$ from the brane at $y=0$ (called the ``Planck
brane''), $a(\yT)$ is exponentially small.  If there was another brane
located at $\yT$, dubbed the ``TeV brane,'' any particles constrained to
exist there would have their masses renormalized by the factor $a(\yT)$. 
Thus even if all mass parameters in the fundamental Lagrangian were of
order $M_P$, physical masses at position $\yT$ could easily be of order
$M_W$ or 1 TeV with only a moderately large value of $\yT$.  The function
$a(y)$ can be interpreted as the wave function of the graviton, showing
that gravity is trapped near the brane at $y=0$. Because of this trapping,
the usual gravitational force law $F\sim 1/r^2$ is obeyed at distances
$r\gg 6M^3/b \Lambda$, even if the extra dimension is infinite in size. 
It is not obvious whether this happy state of affairs is compatible with
getting the correct rate of expansion on the brane at $\yT$, but we shall
show that it is in fact possible to have both. 

An obvious question is whether these ideas can be extended to larger
numbers ($N$) of extra dimensions, since it is possible that
qualitatively new effects might emerge.  Thus far no solutions have
been constructed for a single 3-brane in $N>1$ spatial dimensions.
However it is straightforward to do so for a brane with $(3+N-1)$
spatial dimensions (in other words, with codimension 1).  Moreover, by
taking the intersection of $N$ such branes, one can single out a 3D
region of space which might be identified with a universe like ours,
and this kind of solution has also been constructed, in the static case
\cite{ADDK,angles}.  The static solutions manifest the
phenomenon of gravitational trapping and the potential for solving the
hierarchy problem analogous to the $N=1$ case.  Here we wish to
consider the generalization to dynamical (expanding or contracting)
solutions.  We shall see that an expression similar to
(\ref{rightrate}) obtains for the Hubble rate in the intersecting brane
model.

To specify the solutions, we consider the case of $N$ extra
dimensions with coordinates $y_i$, and $N$ orthogonally intersecting
$(3+N)$-branes located at $y_i=0$, respectively.  The
action is
\beq
	S = \int d^{4}\!x\, d^{N}\!y\, \sqrt{|g|}
	\left({\courbure\over 2\kappa^2} - \Lambda_b - \sum_{i=1}^N \sigma_i
	\delta\left(y_i\sqrt{g_{ii}}\right) \right),
\eeq
where $\kappa^2$ is related to the $N$-dimensional gravity scale $M$
by $\kappa^{-2} = M^{N+2}$.  Similarly to ref.\
\cite{ADDK}, we take the conformally flat ansatz
\beq
	ds^2 = a^{2}(\tau,y_i)
	\left(- d\tau^2 + \sum_{i=1}^3 dx_i^2 + \sum_{j=1}^N dy_j^2\right)
\eeq
for which the Einstein tensor in $d=4+N$ dimensions has the form
\beq
   G_{\mu\nu} = (N+2)\left(a \nabla_\mu \nabla_\nu a^{-1} -
	\eta_{\mu\nu}\left(a\nabla^2 a^{-1} - {N+3\over 2}
	\left(a\nabla a^{-1}\right)^2\right)\right).
\eeq
The gradients are simple partial derivatives, using the Minkowski
metric $\eta_{\mu\nu}= {\rm diag}(-1,1,\dots,1)$: $\nabla^2 = 
\eta^{\mu\nu}\partial_\mu\partial_\nu$.    
If the respective branes have tensions $\sigma_i$,
the stress-energy tensor is given by
\beq
	 T_{\mu\nu} = -a^2\left(\Lambda_b\eta_{\mu\nu} + 
	\sum_{i=1}^N\left(\eta_{\mu\nu} - 
	\delta_{\mu,y_i}\delta_{\nu,y_i}\right)
	\sigma_i 
	\delta(a y_i)\right)
\eeq
Each brane contribution looks like a bulk cosmological term, except 
in the entry corresponding to $y_i$ which is zero for the $i$th brane.
	
A solution to the $(4+N)$ dimensional Einstein equations, $G_{\mu\nu} =
R_{\mu\nu} - \frac12 g_{\mu\nu} \courbure  = \kappa^2 T_{\mu\nu}$,
is given by
\beq
	a(\tau,y_i) = \left(-H\tau + \sum_i k_i|y_i|\right)^{-1}.
\eeq

For the case $N=1$, that is, a single 3-brane, this solution belongs
to a general class of solutions constructed by ref.\ \cite{ChRe} in models that
generalize the RS scenario. This solution was also found (for $N=1$) by
ref.\ \cite{Kim}.

It is easy to show that the equations are satisfied provided that
\beq
	k_i = {\kappa^2 \sigma_i\over 2(N+2)}
\eeq
and the Hubble constant is given by
\beq
	H^2 = {2\kappa^2\Lambda_b\over (N+2)(N+3)} + \sum_i k_i^2.
\eeq
In particular, the static case where $H=0$ is recovered if $\Lambda_b$
satisfies
\beq
\label{statcond}
	\Lambda_b = -{(N+3)\over 8(N+2)} \kappa^2 \sum_i \sigma_i^2.
\eeq
One can see that this agrees with the previous result (\ref{RSrel}) in the
case of one extra dimension.

Our solution differs from previous ones, such as
refs.\ \cite{Nihei,Hatanaka}, by allowing the extra dimensions to
inflate simultaneously with the 3D universe. The inflation of the bulk
causes gravity to become increasingly weaker on the TeV brane, as
discussed below, so this kind of solution is not of direct interest for
late-time cosmology, but might be applicable during an inflationary
phase.

To see that $H$ is indeed the Hubble parameter, one can transform from
the conformal time coordinate $\tau$ to FRW time $t$, in which the
$g_{00}$ element of the metric is $-1$:  $dt = \pm a(\tau,y_i)d\tau$, 
$t = \pm H^{-1}\ln(a)$.  This implies $a(\tau (t),y_i) = \exp(\pm Ht)$.
Choosing the upper sign gives the line element
\beq
	ds^2 = -\left(dt + H^{-1}e^{Ht}\sum_{i=1}^N k_i\, 
	{\rm sign}\,(y_i)\, dy_i
	\right)^2 + e^{2Ht}\left(d\vec x^{\,2} + d\vec y^{\,2}\right)\, .
\eeq
The 4D part of the metric has the usual form for an
inflationary solution with expansion rate $H$.  The range of the $\tau$
coordinate is $\tau\in(-\infty,\sum_i k_i |y_i|/H)$, corresponding to
$t\in(-\infty,+\infty)$.  As $\tau\to 0$ the volume of intersection region 
grows without bound.

To construct a realistic inflationary scenario, one should
replace the constants $\sigma_i$ with time-varying energy densities:
\beq
	\sigma_i \to \sigma_i + V_i(t)
\eeq
Here it is envisioned that the $\sigma_i$ and $\Lambda_b$ satisfy the
condition (\ref{statcond}) which ensures that the expansion will stop
when the perturbations $V_i$ settle to their minimum values, presumed
to be $V_i(\infty)=0$.  The $V_i$ should thus be regarded as potentials
of scalar fields.  The solution we have obtained is not exact for
time-dependent $V_i$'s, but in the limit where they are changing
adiabatically with time, it gives the correct instantaneous rate of
expansion.  Linearizing in these perturbations gives an expansion rate
of
\beq
\label{Hrate}
  H \cong \left(\sum_i {\kappa^4 \sigma_i V_i\over
	2(N+2)^2}\right)^{1/2}.
\eeq
In our approach, $V_i$ represents the energy per unit
$(N+2)$-D volume on the $i$th brane.  
We note that the 3-D intersection of all $N$ branes has a vanishing
$(N+2)$-D volume in the limit of zero brane thickness, $\Delta=0$.
Since the total energy 
density is the sum of the individual brane
contributions, the 3-D energy density in the region of intersection
of all $N$ branes is zero if $\Delta=0$.  
For example in the case $N=2$, the total energy density would be 
proportional to $\delta(y_1)+\delta(y_2)$, which has vanishing support
at the point $y_1=y_2=0$.  To remedy this we must assume that 
$\Delta\neq 0$.  Then the delta functions are replaced
by top-hat functions of width $\Delta$.  
In the $N=2$ case it is clear
that if $V_i$ is the 4-D spatial energy density, then in the intersection
region the 3-D energy density is $(V_1+V_2)\Delta$.  For $N$ extra
dimensions this generalizes to
\beq
	\rho = \sum_i V_i \Delta^{N-1},
\eeq
and the usual rate of expansion, $H = (\rho/3M_P^2)^{1/2}$,
can be obtained by setting 
\beq
\label{sigmacond}
	\sigma_i = { 2(N+2)^2\Delta^{N-1}\over 3\kappa^4 M_P^2}
\eeq 
in agreement with the value (\ref{Lrel}) in the $N=1$ case.  In terms
of the fundamental gravity scale $M$, defined by $\kappa^2 =
M^{-(2+N)}$, it seems reasonable to imagine a brane thickness on the
order of $\Delta \sim M^{-1}$, so that $\sigma_i \sim M^{N+5}/M_P^2$.
This construction leaves unanswered the question of why the matter we
see in our universe, if the latter is the intersection of several
branes, is constrained to stay in that region.  The problem obviously
does not arise in nonintersecting brane scenarios.  For instance, in
the case of one extra dimension compactified on $S^1/{\mathbb{Z}}_{2}$,
the two branes localized on the fixed points do not interact provided
they correspond to different gauge groups.  The standard model resides
on the positive tension brane, whose matter is neutral under the hidden
sector of the other, negative tension brane.  However in the present
proposal, for $N>1$ matter lives in higher (N+2)-dimensional branes.
One is left not only with the question of how the matter which we see
is prevented from moving out of the intersection point in a direction
along one of the branes, but also why it does not seem to interact with
similar matter in the branes but located away from the intersection
point.

We have argued that an observer at the intersection of $N$ branes in
$4+N$ dimensions will experience a rate of cosmological
expansion in accord with the usual Friedmann equation, $H\propto
\sqrt{V_T}$, provided that the conditions (\ref{sigmacond}),
(\ref{statcond}) are satisfied, and that the apparent 3D energy density
is small compared to that coming from the brane tensions, $\rho\ll
\sigma_i\Delta^{N-1}$.  However for this observer there is no immediate
solution to the weak scale hierarchy problem.  Only for a 3-brane which
is located some distance away from the intersection region are masses
suppressed by the geometrical factor $a(y_i)$.  A potential problem is
whether the Hubble rate will be correct when measured on this ``TeV
brane,'' which presumably should have a smaller tension than the
``Planck brane'' intersection region, so as not to significantly
perturb the geometry induced by the Planck brane.  One might expect the
expansion of the universe to be controlled by the large energy density
on the Planck brane, rather than the small one on the TeV brane.  An
observer on the latter might find his universe expanding at a rate that
was not directly correlated with the local energy density.

To investigate this question we will consider the simplest case, that of
$N=1$.  The extension of our previous solution to incorporate a Planck
brane and a TeV brane, having respective positions $y=\yP$, $y=\yT$
and tensions $\sP$, $\sT$, is
\beq
\label{newsoln}
	a(\tau,y) = \left\{ \begin{array}{ll} -H\tau + \kP|y-\yP|, &
	y < \yT \\ -H\tau + \kT|y-\yT| + \kP|\yP-\yT|, &
	y \ge \yT \end{array} \right.
\eeq
This is a generalization of the static solution
found in ref.\ \cite{Hatanaka}.
By computing the $G_{\mu\nu}$ for this metric one finds that it
solves the Einstein equations if 
\beq
	\kP = {\kappa^2\over 6} \sP; \qquad \kT = {\kappa^2\over 6}
	(\sP + 2\sT),
\eeq
and if $\Lambda_b$ changes discontinuously at the interface 
provided by the TeV brane,
\beq
	\LT - \LP \equiv 
	\Lambda_b\Big|^{y=\yT+\epsilon}_{y=\yT-\epsilon}
	= -{2\kappa^2\over 3}\sT(\sT + \sP)
\eeq
The Hubble rate is given by 
\beqa
\label{Hrates}
 H^2 &=& {\kappa^4\sP^2\over 36} + {\kappa^2\LP\over 6} \nonumber\\
     &=& {\kappa^4(\sP+2\sT)^2\over 36} + {\kappa^2\LT\over 6} 
\eeqa

Let us first construct the static configuration where $H=0$.  The term
$H\tau$ can be replaced by a constant in eq.\ (\ref{newsoln}) to maintain
the regularity of the solutions.  The bulk cosmological constants in the
two regions $y<\yT$ and $y>\yT$ are related to the brane tensions by
\beqa
	\LP &=& -{\kappa^2\over 6}\sP^2;\nonumber\\
	\LT &=& -{\kappa^2\over 6}(\sP+2\sT)^2.
\eeqa
To get expanding solutions, we now perturb around the static case by
adding small energy densities $\VP$ and $\VT$ to the branes, and
linearizing.  Eq.\ (\ref{Hrates}) becomes
\beqa
\label{Hrate1}
 H^2 &\cong& {\kappa^4\over 18}\sP\VP \\
\label{Hrate2}
     &\cong& {\kappa^4\over 18}(\sP+2\sT)(\VP+2\VT)
\eeqa
The fact that eqs.\ (\ref{Hrate1}) and (\ref{Hrate2}) must agree implies
that the perturbations on the two branes are proportional, 
\beq
\label{Vrel}
	\VT = -{\sT\over 2\sT + \sP} \VP \cong -{\sT\over \sP} \VP,
\eeq
so that an observer on the TeV brane would relate the expansion rate to
his local energy density by
\beq
\label{Hrate3}
	H^2 = -{\kappa^4\over 18}\,{\sP\over\sT}\,(2\sT+\sP)\,\VT
	\cong -{\kappa^4\over 18}\,{\sP^2\over\sT}\,\VT
\eeq
Now the Planck brane tension $\sP$ must be positive to ensure that
$a(y)$ is decreasing away from $y=\yP$, as is needed to solve the
hierarchy problem; then eq.\ (\ref{Hrate1}) implies $\VP>0$ as well.
From (\ref{Hrate3}) it follows that $\sT$ and $\VT$ must have the
oppposite sign.  This is an improvement over the original Randall-Sundrum
proposal, where the extra dimension being compactified on a circle led
to the topological restriction that $\sT = -\sP$, hence the conclusion
that $\VT$ had to be negative.  In the present realization we can take
$\sT$ negative but smaller in magnitude than $\sP/2$, leading to the 
conclusion that $\VT>0$.  Since we would like $\VT$ to represent the
energy driving cosmological expansion as seen on the TeV brane, this
is encouraging.

In the above construction, we observe that it is not after all
necessary to assume that $\sT\ll \sP$, as might be suggested by the names
``TeV'' and ``Planck'' for the two branes.  All that is really needed
is to have $-\sP/2 < \sT < 0$.  As long as this is true, all the
quantities $\sP$, $\sT$, $\LP$ and $\LT$ can be of order $M_P$ to the
appropriate power.  Then the expansion rate goes like $H\sim
\sqrt{V_i}/M_P$ in terms of the excess energy density on either brane,
as desired.  Moreover it looks straightforward to generalize this
construction to higher dimensions. Then different $N$-dimensional
hypercubic regions would have different values of the bulk cosmological
constant, changing discontinuously at the interfaces where the
analogues of the TeV brane are located.

In the original version \cite{LargeDim} of large extra dimensions,
inflation of the latter was associated with time variation of Newton's
constant, since the largeness of the Planck mass was linked to the size
of the extra dimensions.  It can be seen that inflation of the bulk
actually has no effect on Newton's constant on the Planck brane, but it
does cause the strength of gravity to decrease on the TeV brane.
Ref.\ \cite{ADDK} showed that the relationship between the fundamental
gravity scale and the observed Planck mass is
\beq
\label{MPok}
	M_P^2 = M^{N+2} \int d^{\,N}\!y\, a^{2+N}(y_i)
\eeq
in the static case. This comes from integrating $\sqrt{g}\courbure$ over the
extra dimensions to find the effective 4D action, and using the scaling
property of the Ricci scalar under conformal transformations of the
metric. Applying this to our dynamical solution gives
\beq
\label{MPwrong}
	  M_P^2    = {2^N M^{N+2}\over (N+1)!\prod_i k_i }\, a^2(\tau,0).
\eeq
Recall that $a^2(\tau,0) = (H\tau)^{-2} = \exp(2Ht)$.  Let us now compare 
this to the physical mass of a particle trapped on a brane located
at position $y=\yT$ in the bulk, using the case of $N=1$ extra dimensions
to illustrate.  As first noted by Randall and Sundrum
\cite{RS}, the physical mass ($m_p$) of a particle on such a brane 
is related to the mass parameter in the Lagrangian, $m_0$, by
$m_p = a(\tau,\yT) m_0$.  Therefore the ratio of particle masses on the
TeV brane to the Planck mass scales like
\beq
\label{MPratio}
	{m_p\over M_P} \sim {a(\tau,\yT)\over a(\tau,0)}
	 = {1\over (1 - k|\yT|/H\tau)}
\eeq
which tends to zero as the universe expands ($\tau$ approaches zero
from below).  Translated to FRW time, this says that $m_p/M_P \sim
e^{-Ht} \sim 1/a(t)$.  If such a redshifting of the stength of gravity
were occuring today, it would have been observed by lunar laser ranging
experiments \cite{WND}, which obtain the much more stringent limit
$\dot G/G < (1.25\times 10^{11} {\rm y})^{-1}$.  That is, $\dot G/G$ is
observed to be much less than the present Hubble constant, in
contradiction to the kind of time-dependence given by (\ref{MPratio}).
On the other hand at the Planck brane, $y=0$, (\ref{MPratio}) is constant,
so time variation of Newton's constant would not be observed there.

One possibly annoying feature of our construction is the negative tension
attributed to the TeV brane.  Such a brane might be unstable to
crinkling up into something with an infinite volume since, in the
absence of some stabilizing mechanism, this would minimize the energy.
However in a superstring context such a negative brane could be realized
at an orientifold, which removes the unstable mode \cite{RSstring}.
It is also possible to arrange for positive tension TeV branes; as pointed
out in ref.\ \cite{Oda}, in solutions with several parallel branes in
$N=1$, the signs of the tensions alternate.  Thus one could create a
positive tension TeV brane if desired.  However a negative tension
brane between the TeV and Planck branes is still required, so it is not
clear whether this would be an improvement.

To demonstrate this is a straightforward generalization of our
previous solution.  If we add a third brane at $y=\yTp>\yT$ with
tension $\sTp$, then we have the previous relations, and in addition
\beqa
	\kTp &=& {\kappa^2\over 6}
	(\sP + 2(\sT+\sTp)),\nonumber\\
	\LTp &=& -{\kappa^2\over 6}(\sP+2(\sT+\sTp))^2,
\eeqa
in order to maintain the static condition.  If we now add excess
energies $\VP$, $\VT$ and $\VTp$ to the respective branes, it is 
straightforward to show that in terms of $\VTp$ the expansion rate
is given by 
\beq
	H^2 = -{\kappa^4\over 18}\,{1\over\sTp}\,(2\sT+\sP)\,
	(2(\sT+\sTp) + \sP)\VTp
\eeq
If $\sTp>0$ and $\VTp>0$, then $\sT$ must be in the range
$-\sP/2 - \sTp < \sT < -\sP/2$, showing that the middle
brane has negative tension, although the outer ones have positive
tension.

A final mystery is the question of why the time-dependent parts of the
energy densities of the two branes should be proportional to each
other.  A priori one would think that they are parallel universes which
could have arbitrarily different sources of stress-energy.  Whether
this is an artifact of having a simple ansatz for the solutions, or
there is some deeper reason, is not obvious (but see ``Note Added,''
below).

In summary, we have found expanding global solutions for a $(4+N)$
dimensional universe, in which the intersection of $N$ orthogonal
branes of codimension one plays the role of a 4D universe.  The rate of
expansion can be made to agree with the Friedmann equation if the brane
tensions $\sigma$ are balanced against a negative bulk cosmological
constant $\Lambda_b$ in a particular way.  Both quantities can be of
order $M_P$ to the appropriate power, so only a tuning of their
relative values is necessary; the magnitudes of $\Lambda_b$ and
$\sigma$ are natural.  An energy density $\rho$ which is in excess of
these particular values for the brane tensions is what appears to drive
the expansion of the universe, at the expected rate
$H\sim\sqrt{\rho}/M_P$.  Furthermore it is possible to introduce extra
branes whose physical masses are exponentially suppressed by the
distance from the primary brane, thereby possibly solving the weak
scale hierarchy problem on the extra branes, while maintaining the
correct rate of cosmological expansion.  However, this combination of
two virtues seems to come always at the expense of introducing some
negative tension branes.  Our solutions are not suitable for late-time
cosmology on the TeV brane because the inflating bulk causes gravity to
decouple there.  On the Planck brane this is not a problem, but
new long-range forces due to exchange of the massless excitation
associated with the expansion of the bulk would be \cite{GW}.

{\bf Note Added:}  After this paper was completed, ref.\ \cite{CGRT} observed
that the fine-tuned relationship between energy densities on the two
branes is always a consequence of demanding a static bulk, even in the
absence of a mechanism for stabilizing the bulk.  It is interesting to
note that such a relation is also required in our solutions, even though
the bulk is not stable, but inflating.  In our case, the relationship must
therefore come from the coincidence that the bulk is inflating at exactly
the same rate as the branes. 

\section*{Acknowledgements}

C.G. is grateful to the financial support of the Service de Physique Th\'eorique,
CEA Saclay where this work has been initiated.
This work was supported in part by the Director, Office
of Energy Research, Office of High Energy and Nuclear Physics, Division of
High Energy Physics of the U.S. Department of Energy under Contract
DE-AC03-76SF00098 and in part by the National Science Foundation under
grants PHY-95-14797.


\end{document}